\documentstyle[aps,floats,epsf]{revtex}

\tightenlines

\begin{document}
\draft
\title{Structural Properties of Two-Dimensional Polymers
\footnote{to appear in Physica A}}
\author{Christian M\"unkel and Dieter W. Heermann}
\address{
        Institut f\"ur Theoretische Physik \\
        Universit\"at Heidelberg \\
	Philosophenweg 19\\
	6900 Heidelberg\\
	and\\
	Interdisziplin\"ares Zentrum\\
	f\"ur wissenschaftliches Rechnen\\
	der Universit\"at Heidelberg\\
        Germany
     }
\date{May 12, 1993}

\maketitle

\begin{abstract}
We present structural properties of two-dimensional polymers as far as
they can be described by percolation theory. The percolation threshold,
critical exponents and fractal dimensions of clusters are determined by
computer simulation and compared to the results of percolation theory.
We also describe the dependence of the typical cluster structures on the
reaction rate.
\end{abstract}


\vfill\eject

\twocolumn

\section{Introduction}
\par Polymers are known to develop various forms,  for example
linear chains, stars, rings, combs, ladders, three-dimensional networks
and branched polymers. 
All these kinds of polymers 
have linear chains between the branching point.
The kind of two-dimensional polymers we are studying are compact sheets of 
monomers, which are linked
together periodically with respect to two dimensions. Between the branching
points there is no chain and the polymers are 
highly organized sheets.
\par The successful ``in bulk'' synthesis of such two-dimensional polymers
was reported recently \cite{stupp93,thomas93}.
Stupp et al. polymerized self-organized bilayers by two
different so-called stitching reactions, which act within three distinct
levels of the bilayer. Each oligomer can not
have more than two bonds within a layer. In addition, not all possible
bonds are present, because the stitching reactivity is about 30 to 50\% at
the upper and lower layer (90\% within
the middle layer). Is it possible to form a two-dimensional polymer
instead of an ensemble of ladder polymers? What is the minimal reaction rate,
needed for large two-dimensional polymers? What kind of structures will be
developed at different reaction rates? 
\par These are questions, which can be answered by percolation theory
\cite{flory41}.

\section{Results for the stitching reaction}
\par One of the most important results of percolation theory
\cite{stauffer92,binder92,heermann90} is related to ``universality'',
if the system has short ranged interactions.
All presently available
evidence strongly suggests, that the critical exponents depend on the
dimensionality of the lattice only --- but not on the lattice structure,
boundary conditions and so on.
The exponents for bond or site percolation, for square, triangular or
honeycomb lattices etc. are the same. We therefore would expect, that
the clusters produced with the above cited stitching reaction
also may be described by these
exponents.

\par Does there exist a large (``infinite'') cluster of bonded sites?
To answer this question,
we define the fraction of sites in the largest cluster $P_\infty$ and the
probability $P_S$, that there is a ``spanning cluster'' of
bonds, which connects two opposite boundarys of the underlying lattice.
Then for an infinite lattice, $P_\infty$ is expected to vanish for
$(p-p_c) \rightarrow 0_+$ as
\begin{equation}
	P_\infty(p) \propto {\left(p - p_c\right)}^{\beta} \; \; \; ,
\label{equ:def_beta}
\end{equation}
where $p$ is the probability, that a bond is present, and $p_c$ is the
percolation threshold, above which an infinite cluster exists.
With $n_s$ clusters of size $s$ (number of polymers)
the mean size of the finite
clusters is related to the percolation susceptibility
$\chi$ (omnitting the largest (``infinite'') cluster)
\begin{equation}
\chi \approx  {\sum_{s=1}^{\infty}}' \, s^2 \, n_s(p) \; \; \; ,
\label{equ:def_chi}
\end{equation}
which diverges at $p_c$ as
\begin{equation}
	\chi(p) \propto {\left|p - p_c\right|}^{-\gamma} \; \; \; .
\label{equ:def_gamma}
\end{equation}

Finally, at $p=p_c$ the number $n_s$ of clusters with size $s$ 
is expected to decay with a power of $s$:
\begin{equation}
	n_s(p_c) \propto {s}^{-\tau}
\label{equ:def_tau}
\end{equation}
Here $\beta, \gamma \mbox{and} \tau$ are the mentioned exponents.

\par Stupp et al. \cite{stupp93} suggest, that they synthesized large
two-dimensional polymers. Obviously, the effective reaction rate of the
stitching reaction has to be larger than the percolation threshold $p_c$,
if one wants to synthesize large sheets instead of ladder polymers.
Since $p_c$ is not universal, we use a simple but sufficient model
of the two-dimensional polymer, to 
compute $p_c$. In addition we will measure the exponents.

\par
\begin{figure*}[htbp]
\epsfysize=17cm
\centerline{\epsffile{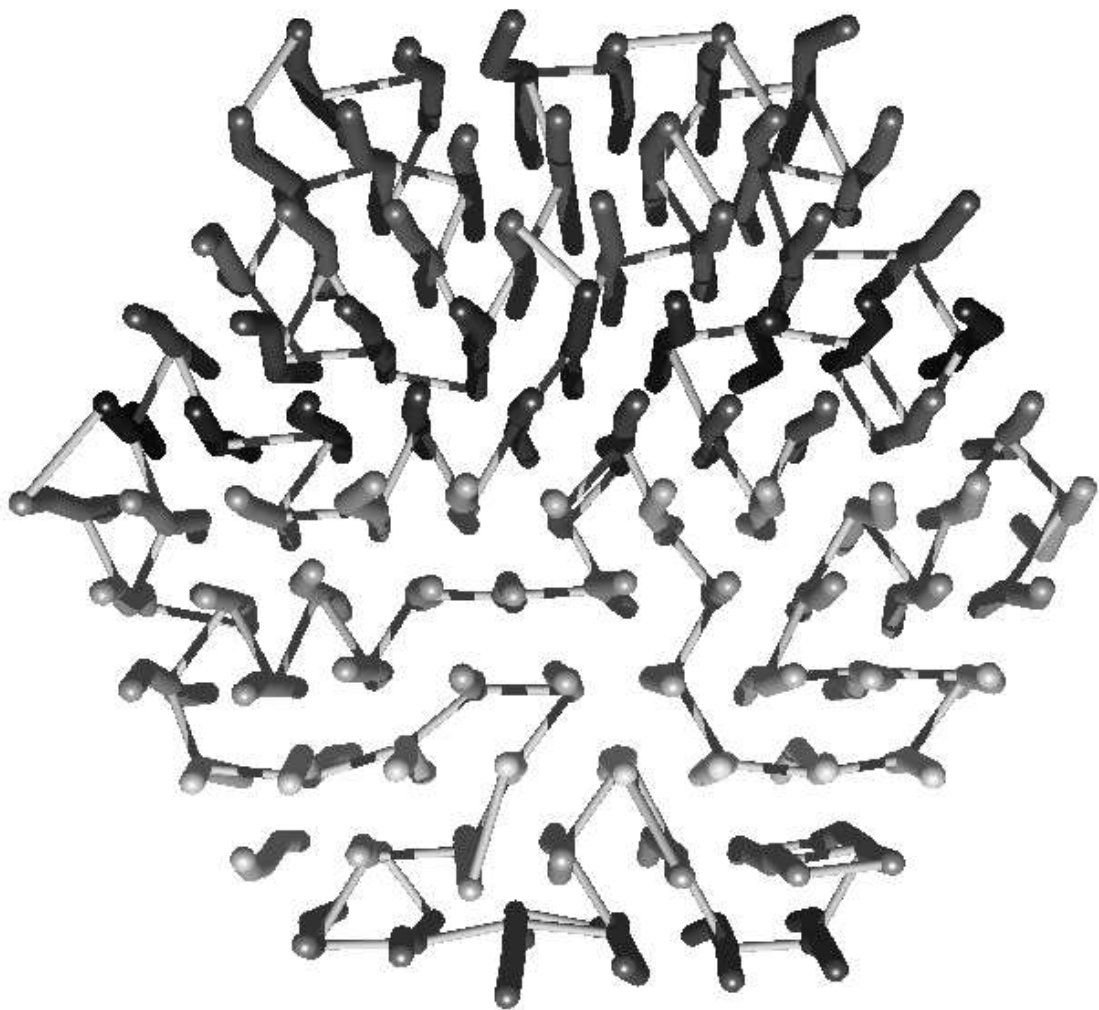}}
\caption{\em Schematic representation of the two-dimensional polymer model
	of linear size $L=11$ and $p=0.4$.
         }
\label{fig:schematic}
\end{figure*}

 As a model of the two-dimensional polymer we used a stack of
three layers, each triangular, hexagonally shaped with free boundaries,
as shown in figure~\ref{fig:schematic}.
At the middle layer, the reaction rate was 100\%, i.e. all allowed bonds (two
can react) were occupied. For that purpose, the bonds of the triangular lattice
were choosen successively at random. If the number of bonds of each polymer was
less than 2, the randomly choosen bond was occupied. Then we stitched the
upper and lower layer using the same algorithm, until $n_{+1}$ ($n_{-1}$) bonds
were occupied. $n_{+1}$ ($n_{-1}$) is the number of occupied bonds
of the middle layer $n_0$ times the stitching probability (reaction rate)
$p$ of the upper and lower layer.
Note, that $n_0$ is not determined by the linear lattice size $L$, but may
vary slightly for each configuration.
\par Because of the restriction, that each polymer may be stitched to at most
two neighbours, we do not expect an infinite cluster or percolation at
$p=0$, i.e. without bonds at the upper and lower layer, but an ensemble of
linear polymer chains. On the contrary, at $p=1$ all allowed bonds at all
three layers are occupied, resulting in an effective coordination number larger
than two. It is worth noting,
that even at $p=1$ the probability, that there is only one cluster containing
all polymers, vanishes with increasing system size $L$. Within this model,
it is very unlikely to prepare very large perfect two-dimensional
polymers.
\par
\begin{figure*}[htbp]
\epsfxsize=12cm
\centerline{\epsffile{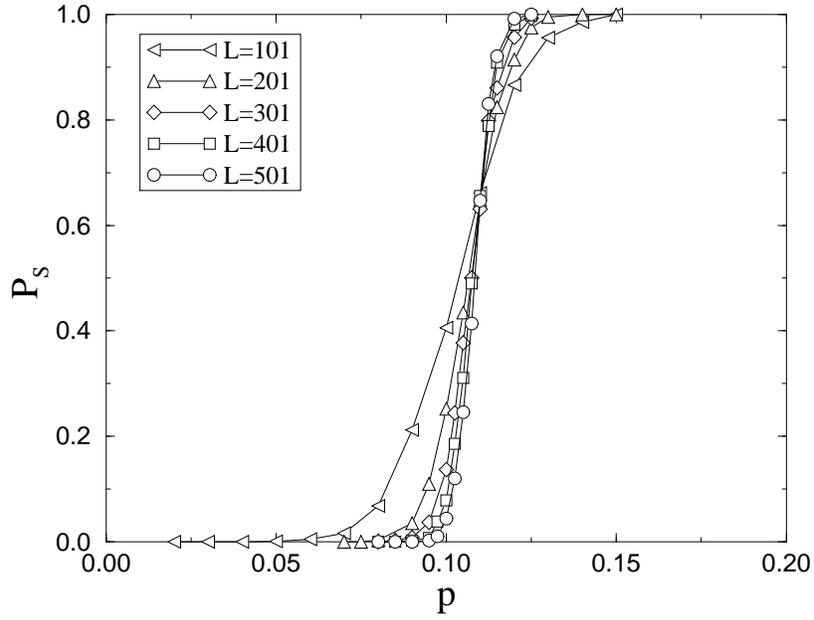}}
\caption{\em Unbiased estimate of the percolation threshold $p_c = 0.110(3)$
	using the spanning probability $P_S$.
         }
\label{fig:fss_P_S}
\end{figure*}

 If we want to answer the question, if a reaction rate of about $40$\%
is high enough to prepare infinite clusters, we have to determine the
percolation threshold. As noted,
the value of $p_c$ is not universal, but depends strongly on the used lattice
and bond restrictions. An unbiased way to estimate $p_c$ is to plot the
probability $P_S$ of the occurence of a spanning cluster as a function of 
$p$ for different lattice sizes \cite{binder92}. The finite size scaling
relation for $P_S$ simply reads
\begin{equation}
	P_S^{(L)} (p) = \tilde{P_S}\left({\frac{L}{\xi_p}}\right)
			 \; \; \; , \nonumber
\end{equation}
where $\tilde{P_S}$ is the scaling function and $\xi_p$ the correlation
length. As a consequence, different curves $P_S^{(L)}(p)$ for different
choices of $L$ should intersect at $p=p_c$ in a common intersection point
$\tilde{P_S}(0)$, apart from corrections to scaling and a bias by the
choice of the critical exponents. Figure~\ref{fig:fss_P_S} shows this plot
for the stitching reaction described above. 
\par From this figure we get one of our main result, the percolation threshold
$p_c = 0.110(3)$. Hence it follows, that
the reaction rate of 30\% to 50\% reported by Stupp et al.
\cite{stupp93} was high enough
(within the limits of our percolation model) to prepare an ``infinite''
two-dimensional polymer.

\par
\begin{figure*}[htbp]
\epsfxsize=12cm
\centerline{\epsffile{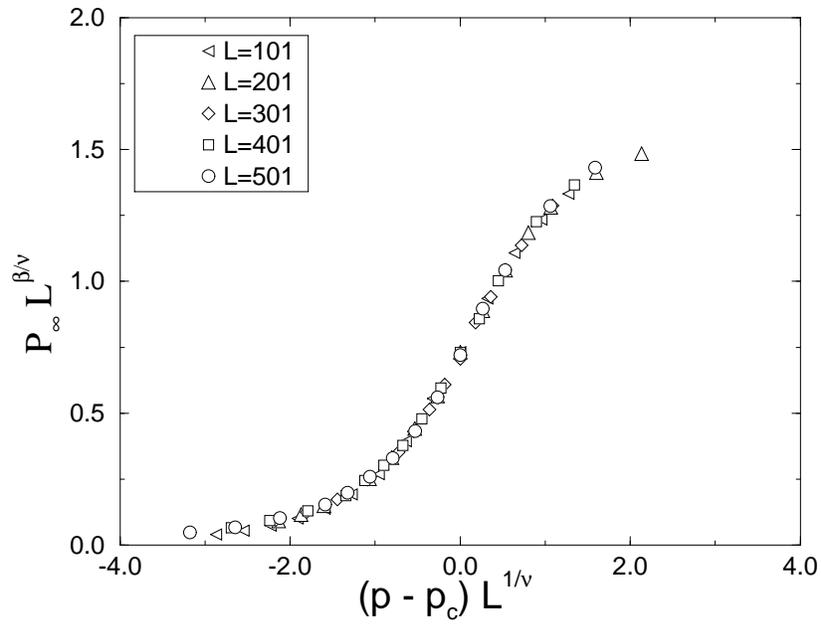}}
\caption{\em Finite size scaling of the probability $P_\infty^{(L)}$, which
	was defined as the fraction of polymers in the largest cluster, 
	with exponents $\nu = 4/3$, $\beta = 5/36$ and $p_c = 0.110(3)$.
         }
\label{fig:fss_m}
\end{figure*}

 Does the percolation of the stitching reaction lie in the same 
universality class as the standard bond percolation? 
Figure~\ref{fig:fss_m} shows a finite size scaling plot of $P_\infty$
using the assumed exact values for $\beta=5/36$ and $\nu=4/3$
and $p_c$ obtained from the spanning probability.
Within the accuracy of the data,
the finite size scaling assumption seems to be fulfilled.

\par
\begin{figure*}[htbp]
\epsfxsize=12cm
\centerline{\epsffile{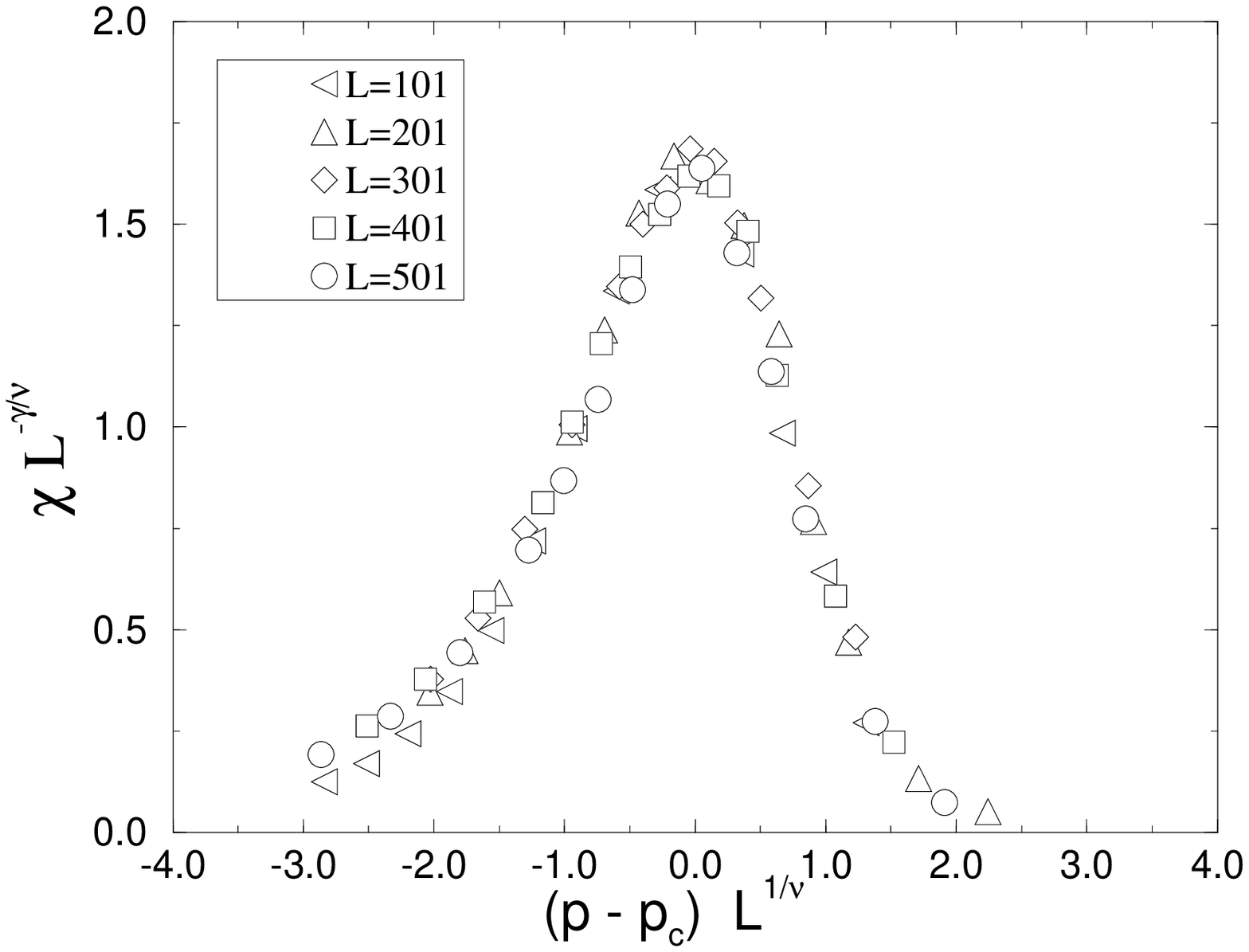}}
\caption{\em Finite size scaling of the susceptibility $\chi^{(L)}$
	with exponents $\nu = 4/3$, $\gamma = 43/18$ and $p_c = 0.110(3)$.
         }
\label{fig:fss_chi}
\end{figure*}

 We may also compare the cluster size distribution to bond percolation.
Using $\gamma=43/18$, we plotted the scaling function
of the susceptibility $\chi$ in figure~\ref{fig:fss_chi}.
Apart from small
deviations of the smallest system far from $p_c$, the system sizes simulated
seem to be in the scaling regime. The maximal values of $\chi$ vary by
orders of magnitude in this range of $L$!

\par
\begin{figure*}[htbp]
\epsfxsize=12cm
\centerline{\epsffile{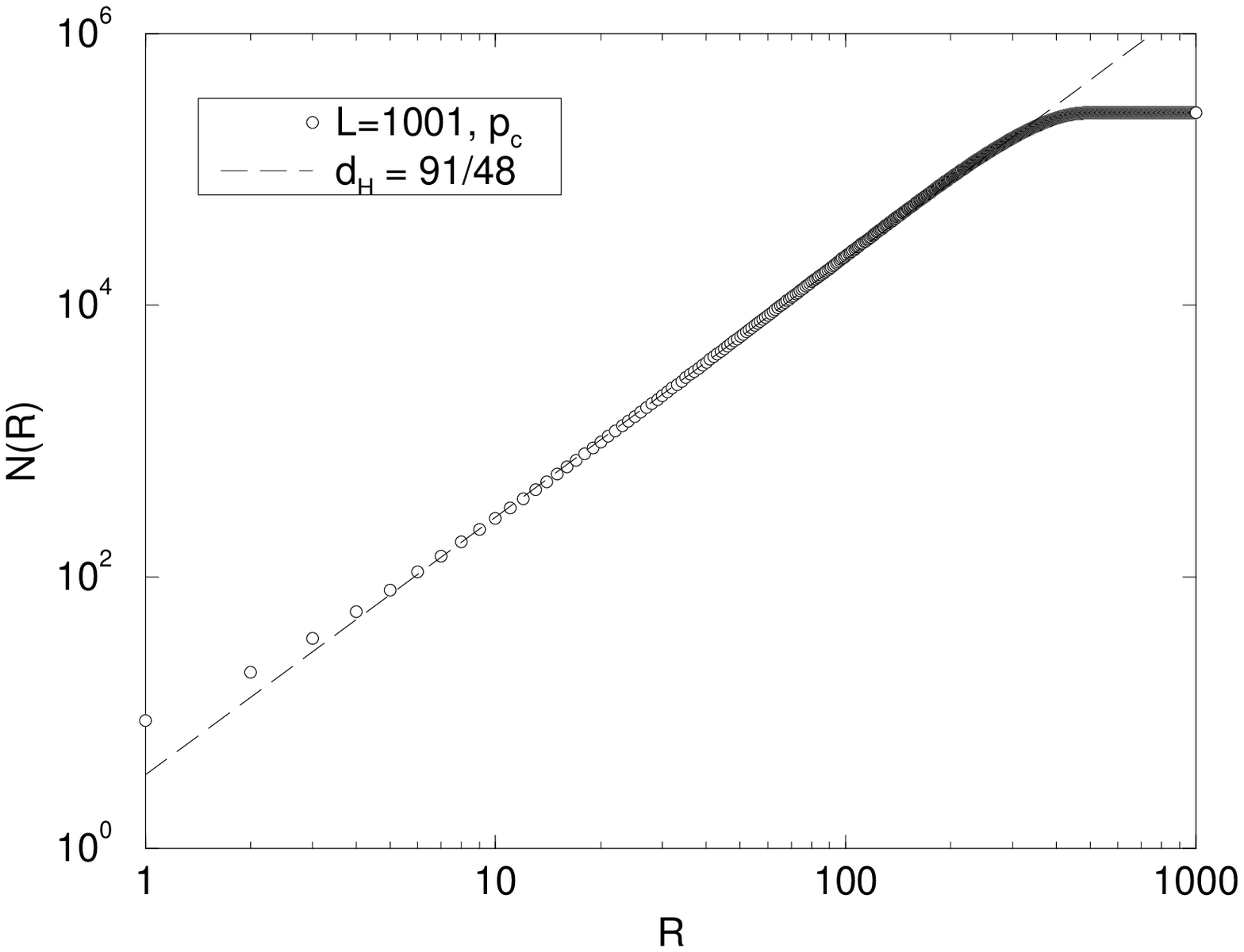}}
\caption{\em Number of polymers $N$ within a radius $R$ of the largest cluster.
The data of $1000$ configurations of $L=1001$ hexagons were accumulated.
	The dashed line has the exact 
	Hausdorff dimension (slope) $d_H=91/48$.
         }
\label{fig:d_H}
\end{figure*}

 Finally, we are interested in the geometric properties of the cluster
structure, particularly at the percolation threshold $p_c$. At $p_c$ the
largest (spanning) cluster should form a fractal with Hausdorff dimension
$d_H$. This quantity can be defined with the number of polymers $N$ within
a sphere of radius $R$ and the relation 
\begin{equation}
	N = \mbox{const.} \times R^{d_H}  \; \; \;  .
\label{equ:def_d_H}
\end{equation} 
Using the data of $1000$ configurations of a $L=1001$ hexagon at $p_c$,
which is shown in figure~\ref{fig:d_H}, we measured $d_H(p_c) = 1.91(5)$
in good agreement with the result of percolation theory $91/48\approx1.896$.

\par
\begin{figure*}[htbp]
\epsfxsize=12cm
\centerline{\epsffile{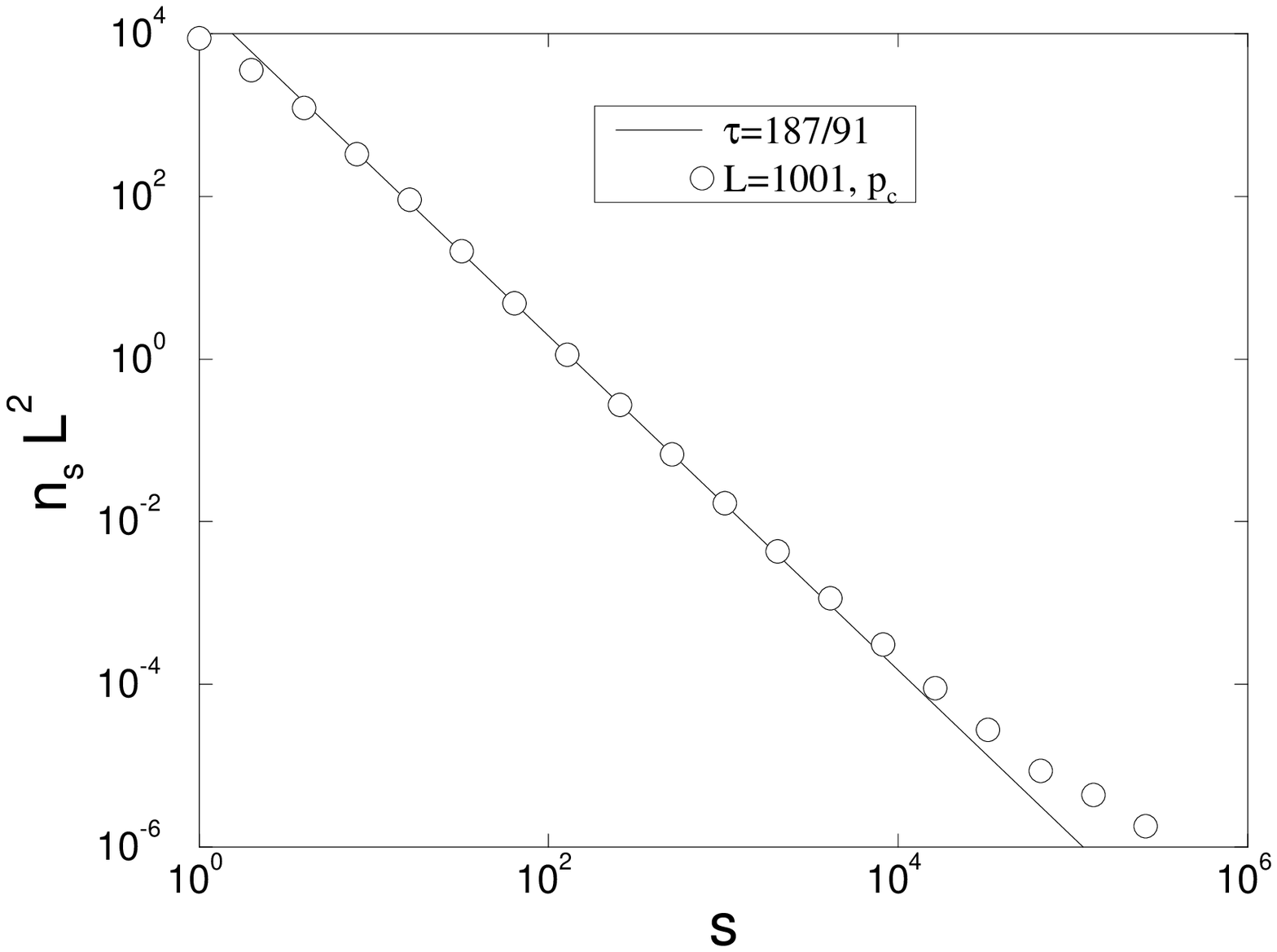}}
\caption{\em Cluster distribution of $1000$ configurations of $L=1001$ hexagon 
($751501$ polymers) at the percolation threshold $p_c\approx0.110$. From the
	slope one computes the Fisher exponents $\tau\approx 1.95(10)$.
         }
\label{fig:cluhis}
\end{figure*}

 Apart from the largest cluster, there are many smaller ones. Because
of the statistical nature of their structure, we restrict ourselves to
the number $n_s$ of clusters of size $s$ (which is the number of polymers),
which is expected to decrease as a power of $s$. Within the accuracy of the
data, the power law seems to be fulfilled in figure~\ref{fig:cluhis}.
Although the Fisher epxonent
$\tau \approx 1.95(10)$ is somewhat smaller than the expected one $187/91$.
Most probably this is due to the finiteness of the system. High--quality
data \cite{rapaport91} showed, that $n_s$ is less than the
expected power law for small $s$ and higher for the largest $s$. Therefore,
we underestimated $\tau$ systematically.


\begin{thebibliography}{1}

\bibitem{stupp93}
S. Stupp, S. Son, H. Lin, and L. Li, Science {\bf 259},  59  (1993).

\bibitem{thomas93}
E. L. Thomas, Science {\bf 259},  43  (1993).

\bibitem{flory41}
P. J. Flory, J. Am. Chem. Soc. {\bf 63}, 3091 (1941).

\bibitem{stauffer92}
D. Stauffer and A. Aharony, {\em Introduction to Percolation Theory, 2nd ed.}
  (Taylor and Francis, 1992).

\bibitem{binder92}
K. Binder and D. W. Heermann, {\em Monte Carlo Simulation in
 Statistical Physics,
  2nd ed.} (Springer-Verlag Heidelberg, 1992).

\bibitem{heermann90}
D. W. Heermann, {\em Computer Simulation Methods in Theoretical Physics,
 2nd ed.}
  (Springer Verlag, Heidelberg, 1990).

\bibitem{rapaport91}
D. C. Rapaport, J. Stat. Phys. {\bf 66}, 1/2, (1992).

\end{thebibliography}
\end{document}